\PassOptionsToPackage{numbers,sort&compress}{natbib}
\documentclass[preprintnumbers,amsmath,amssymb,prd,notitlepage,nofootinbib,superscriptaddress,twocolumn]{revtex4}
\usepackage[utf8]{inputenc}

\makeatletter
\renewcommand{\p@subsection}{}
\renewcommand{\p@subsubsection}{}
\makeatother

\usepackage[numbers,sort&compress]{natbib}
\usepackage[normalem]{ulem}
\usepackage{graphicx}
\usepackage{xcolor}
\usepackage{dcolumn}
\usepackage{bm}
\usepackage{subfigure}
\usepackage{wasysym}
\usepackage{verbatim}
\usepackage{color}
\usepackage{amsmath}
\usepackage[utf8]{inputenc}
\usepackage[hidelinks]{hyperref}
\usepackage{aas_macros}
\usepackage{cprotect}

\newcommand{\N}{\mathcal{N}}

\newcommand{\be}{\begin{equation}}
\newcommand{\ee}{\end{equation}}
\newcommand{\ba}{\begin{eqnarray}}
\newcommand{\ea}{\end{eqnarray}}

\newcommand{\barr}{\begin{array}}
\newcommand{\earr}{\end{array}}

\def\RDN0{\mbox{RDN}^{(0)}}

\def\N32{N^{(3/2)}}

\usepackage{simplewick}

\begin{document}

\title{The Bias to Cosmic Microwave Background Lensing Reconstruction from the Kinematic Sunyaev-Zel'dovich Effect at Reionization}
\author{Hongbo Cai}
 \email{hoc34@pitt.edu}
 \affiliation{Department of Physics and Astronomy, University of Pittsburgh, Pittsburgh, PA, USA 15260}

\author{Mathew S. Madhavacheril}
 \affiliation{Perimeter Institute for Theoretical Physics,
31 Caroline Street N, Waterloo ON N2L 2Y5 Canada}
\affiliation{Department of Physics and Astronomy, University of Southern California, Los Angeles, CA, 90007, USA}

\author{J. Colin Hill}
\affiliation{Department of Physics, Columbia University, New York, NY 10027, USA}
\affiliation{Center for Computational Astrophysics, Flatiron Institute, New York, NY 10010, USA}

 \author{Arthur Kosowsky}
  \affiliation{Department of Physics and Astronomy, University of Pittsburgh, Pittsburgh, PA, USA 15260}
\date{\today}

\begin{abstract}
The power spectrum of reconstructed cosmic microwave background (CMB) lensing maps is a powerful tool for constraints on cosmological parameters like the sum of the neutrino masses and the dark energy equation of state.  One possible complication is the kinematic Sunyaev-Zel'dovich (kSZ) effect, due to the scattering of CMB photons by moving electrons, which can bias the reconstruction of the CMB lensing power spectrum through both kSZ-lensing correlations and the non-Gaussianity of the kSZ temperature anisotropies. We investigate for the first time the bias to CMB lensing reconstruction from temperature anisotropies due to the reionization-induced kSZ signal and show that it is negligible for both ongoing and upcoming experiments based on current numerical simulations of reionization.  We also revisit the bias induced by the late-time kSZ field, using more recent kSZ simulations.  We find that it is potentially twice as large as found in earlier studies, reaching values as large as several percent of the CMB lensing power spectrum signal, indicating that this bias will have to be mitigated in upcoming data analyses.

\end{abstract}
\maketitle
\section{INTRODUCTION}
\label{sec:introduction}
Cosmic microwave background (CMB) photons are deflected by gravitational potentials while they travel from the last scattering surface to us. This effect, known as CMB lensing \cite{lewis06_weak_gravit_lensin_cmb, hanson11_cmb_temper_lensin_power_recon}, remaps the CMB intensity and polarization fields. Compared with galaxy lensing \cite{PhysRevD.92.063517}, CMB lensing can probe matter inhomogeneities at relatively higher redshifts. In addition, CMB lensing also helps constrain fundamental cosmological parameters \cite{Bianchini:2019vxp}, including the sum of the neutrino masses \cite{Santos:2013gqa,2015PhRvD..92l3535A,Bianchini:2019vxp} and the dark energy equation of state \cite{Santos:2013gqa}.

CMB lensing can be described by the lensing potential, which is a line-of-light projection of the three-dimensional gravitational potential weighted by a geometric lensing kernel.  The lensing potential can be reconstructed using a minimum-variance quadratic estimator \cite{Hu:2001kj,Maniyar2021} using both CMB temperature and polarization maps. Ongoing and upcoming experiments, including Advanced ACT \cite{Henderson:2015nzj}, SPT-3G \cite{Benson:2014qhw}, and Simons Observatory (SO) \cite{Ade:2018sbj}, have sensitivities for which temperature maps provide most of the statistical weight for lensing reconstruction; proposed future experiments such as CMB-S4 \cite{Abazajian:2019eic} and CMB-HD \cite{Sehgal:2019ewc} will have high enough sensitivity so that polarization becomes the dominant channel for lensing reconstruction.

Besides lensing, CMB temperature maps also contain signals from several
secondary anisotropies including the thermal and kinematic Sunyaev-Zel'dovich effects (tSZ and kSZ), the integrated Sachs-Wolfe effect, and the Rees-Sciama effect, and foregrounds from thermal dust, radio synchrotron emission, and the cosmic infrared background (CIB). Most of these contaminants can be removed or suppressed by multifrequency component separation methods \cite{Bennett:2003ca, Madhavacheril:2019nfz, Abylkairov2021, Bleem2021}. Since the kSZ effect preserves the blackbody spectrum of the CMB \footnote{There are higher-order relativistic SZ terms which have the same velocity dependence as the kSZ effect \cite{Sazonov:1998ae, Dolgov:2000td, Coulton:2019ign}, and they do not preserve blackbody spectrum.}, it cannot be removed by multifrequency component separation methods (internal linear combination, or ILC) (e.g., as possible for the tSZ or CIB fields~\cite{1802.08230})\footnote{The standard ILC can induce significant bias, which can be mitigated by geometric methods such as profile hardening,  and by partial joint deprojection method \cite{Sailer:2021vpm}. The best-performed combination of multifrequency and geometric methods is introduced in \cite{Darwish:2021ycf}.}. Any significant bias to CMB lensing reconstruction from the kSZ effect must be mitigated using geometric techniques such as
shear-only reconstruction \cite{Schaan:2018tup} or bias-hardened estimators \cite{Namikawa_2013, Osborne:2013nna, Sailer:2020lal}.

The kSZ signal has two physically distinct pieces: the late-time contribution (below the redshift at which the
universe has become fully reionized) and the reionization contribution (from the epoch before reionization is complete). The bias to CMB lensing power spectrum measurements from temperature anisotropies due to the late-time kSZ signal was investigated in \cite{ferraro18_bias_to_cmb_lensin_recon}, showing that the large-scale ($L<500$) fractional bias can reach \(0.5\%\), \(2\%\) and \(3\%\) for Planck, Stage-III experiments (similar to SO) and CMB-S4, respectively, when using  \(\ell_{\mathrm{max}}=4000\), and about half of that for \(\ell_{\mathrm{max}}=3000\), where ${\ell}_{\mathrm{max}}$ is the maximum temperature multipole used in the lensing reconstruction and $L$ is the multipole of the reconstructed lensing map. In this work, we investigate the bias due to the reionization kSZ signal for the first time, and also revisit the late-time kSZ bias. Since the kSZ effect produces negligible polarization fluctuations, we consider only temperature anisotropies. The redshift of reionization and its detailed kSZ signature are not constrained to high precision by current data; our conclusions are based on current numerical simulations of reionization from the WebSky\footnote{\url{https://lambda.gsfc.nasa.gov/simulation/tb_mocks_data.cfm}} extragalactic CMB simulations \cite{Stein_2020}.
Our conclusions will have some quantitative dependence on the exact reionization model assumed, although they are unlikely to
differ qualitatively. In addition, a related but different analysis is performed in \cite{Mishra:2019qyd}, which computes the bias to CMB lensing reconstruction arising from the fact that foreground fields (including the reionization kSZ signal) are lensed by the some of the same structures as the CMB.  The bias arising from the intrinsic non-Gaussianity of the kSZ field is not considered in \cite{Mishra:2019qyd}, and forms the focus of our investigation.

In this paper, we assume a \(\mathrm{\Lambda CDM}\) fiducial cosmology with the Planck 2015 parameters given in the third column of Table 4 of \cite{Ade:2015xua}.

This paper is organized as follows.  We revisit the CMB lensing reconstruction using CMB temperature anisotropies in Section~\ref{sec:lensing reconstrucion}. In Section~\ref{sec:kSZ bias}, we introduce the kSZ effect and explain how to estimate the bias induced by the kSZ effect on the reconstructed CMB lensing power spectrum. We describe the details of our simulations and the numerical results in Section \ref{sec:simulations}, and conclude in Section \ref{sec:discussion and conclusions}.  Analytical details of the CMB temperature trispectrum and the kSZ effect can be found in Appendices~\ref{sec:appendix A} and \ref{sec:appendix B}.

\section{CMB LENSING RECONSTRUCTION FROM TEMPERATURE ANISOTROPIES}
\label{sec:lensing reconstrucion}
A gravitationally lensed CMB temperature map \(\tilde{T}(\hat{\bf{n}})\) can be expressed as
\begin{equation}
  \label{eq:defection}
  \tilde{T}(\hat{\mathbf{n}})=T(\hat{\mathbf{n}}+\mathbf{d}({\hat{\mathbf{n}}})),
\end{equation}
where \(\hat{\bf{n}}\) is the direction on the full sky, \(T(\hat{\bf{n}})\) is the unlensed CMB temperature map, and \(\bf{d}(\hat{\bf{n}})\) is the lensing deflection field. At lowest order in the deflection field, \(\bf{d}(\hat{\bf{n}})\) is a pure gradient \cite{lewis06_weak_gravit_lensin_cmb} given by
\begin{equation}
  \label{eq:deflection}
  \bf{d}(\hat{\bf{n}}) = \nabla \phi(\hat{\bf{n}}),
\end{equation}
where \(\nabla\) represents the angular derivative on the sphere defined by \(\hat{\bf{n}}\), and \(\phi(\hat{\bf{n}})\) is the lensing potential. The CMB lensing convergence is defined as
\begin{equation}
  \label{eq:lensing convergence}
  \kappa(\hat{\bf{n}}) = -\frac{1}{2}\nabla^2 \phi(\hat{\bf{n}}),
\end{equation}
which can be reconstructed by a minimum variance quadratic estimator.  Under the flat-sky approximation in Fourier space, the temperature quadratic estimator is given by \cite{Hu_2002}
\begin{equation}
  \label{eq:kappa estimator}
  \hat\kappa({\mathbf{L}}) = \frac{1}{2}L^{2}A(\mathbf{L})\int_{\bm{\ell}}g(\bm{\ell},\mathbf{L})T^{\mathrm{tot}}(\bm{\ell})T^{\mathrm{tot}}(\mathbf{L}-\bm{\ell}),
\end{equation}
where
\begin{equation}
  \label{eq:weight}
  g(\bm{\ell},\mathbf{L}) = \frac{C_{\ell}^{TT}\bm{\ell} \cdot \mathbf{L}+ C_{|\mathbf{L}-\bm{\ell}|}^{TT}\mathbf{L}\cdot (\mathbf{L}-\bm{\ell})}{2C_{\ell}^{tot}C_{|\mathbf{L}-\bm{\ell}|}^{tot}},
\end{equation}
\(\hat\kappa({\mathbf{L}})\) represents the estimator for \(\kappa(\mathbf{L})\), and $T^{\mathrm{tot}}$ is the observed total temperature anisotropy field containing the lensed primordial fluctuations $\tilde{T}$, the blackbody secondary fluctuations, and  detector noise (which we assume is uncorrelated with the other components). The normalization \(A(\mathbf{L})\) and weight \(g(\bm{\ell},\mathbf{L})\) both depend on the fiducial lensed CMB power spectrum and the experiment's beam and noise properties. The \(C_{\ell}^{tot}\) in the denominator is the total observed CMB temperature power spectrum. Note here we denote
\begin{equation}
  \label{eq:denote}
  \int_{\bm{\ell}} \equiv \int\frac{d^2\bm{\ell}}{(2\pi)^2}.
\end{equation}
The expectation value of the power spectrum of ${\hat{\kappa}}$ is
\begin{equation}
  \label{eq:naive ps}
\langle C_{L}^{\hat{\kappa}\hat{\kappa}} \rangle = \langle\hat{\kappa}(\mathbf{L}) \hat{\kappa}^{*}(\mathbf{L})\rangle = \langle\hat{\kappa}(\mathbf{L}) \hat{\kappa}(\mathbf{-L})\rangle,
\end{equation}
where $\langle \ \rangle$ represents ${\langle \langle \ \rangle_{\mathrm{CMB}} \rangle}_{\mathrm{LSS}}$ which denotes an ensemble average over different primordial CMB Gaussian realizations and large-scale structure realizations \cite{Kesden:2003cc}. Thus, all the maps inside the brackets are random variables. In the real simulation introduced in Section \ref{sec:simulations}, the $C_{L}^{\hat{\kappa}\hat{\kappa}}$ are bandpowers binned on the two-dimensional reconstructed $\kappa$ Fourier map.

To get an unbiased estimator of $C_{L}^{\kappa \kappa}$, several reconstruction biases are subtracted from $C_{L}^{\hat{\kappa}\hat{\kappa}}$ as
\begin{equation}
  \label{eq:noise substraction}
    \begin{aligned}
      \hat{C}_{L}^{\kappa\kappa}  = &\langle C_{L}^{\hat{\kappa}\hat{\kappa}}\rangle - (\Delta C_{L}^{\kappa \kappa})_{\mathrm{Gauss}} - (\Delta C_{L}^{\kappa \kappa})_{\mathrm{N1}}\\- &(\Delta C_{L}^{\kappa \kappa})_{\mathrm{MC}} - (\Delta C_{L}^{\kappa \kappa})_{\mathrm{FG}},
      \end{aligned}
\end{equation}
where $\hat{C}_{L}^{\kappa\kappa}$ is the estimator of $C_{L}^{\kappa \kappa}$, $(\Delta C_{L}^{\kappa \kappa})_{\mathrm{Gauss}}$
is the Gaussian bias induced by the Gaussian disconnected component of the trispectrum in Eq.~\eqref{eq:kappa ps},\footnote{In real experiments, we calculate the realization-dependent Gaussian bias $(\Delta C_{L}^{\kappa \kappa})_{\mathrm{RDN0}}$ which depends on the CMB realizations \cite{Namikawa_2013, Madhavacheril:2020ido}.} $(\Delta C_{L}^{\kappa \kappa})_{\mathrm{N1}}$ arises from the connected terms of the CMB trispectrum containing an integral over \(C_{L}^{\kappa \kappa}\) \cite{Kesden:2003cc}, $(\Delta C_{L}^{\kappa \kappa})_{\mathrm{MC}}$ is a ``Monte Carlo'' (MC) bias that encapsulates biases which have not been accounted for otherwise, such as higher-order corrections\footnote{In this paper, the lensing potential we consider is Gaussian, so no $(\Delta C_{L}^{\kappa \kappa})_{\mathrm{N3/2}}$ appears~\cite{Bohm:2016gzt,Bohm:2018omn,Fabbian2018}.}, and $(\Delta C_{L}^{\kappa \kappa})_{\mathrm{FG}}$ is the foreground bias. In this paper, $(\Delta C_{L}^{\kappa \kappa})_{\mathrm{FG}}$ is limited to the bias from the kSZ effect $(\Delta C_{L}^{\kappa \kappa})_{\mathrm{kSZ}}$.

\section{CMB lensing power spectrum bias from the kSZ effect}
\label{sec:kSZ bias}
\begin{figure}[t]
\includegraphics[width=1.0\linewidth]{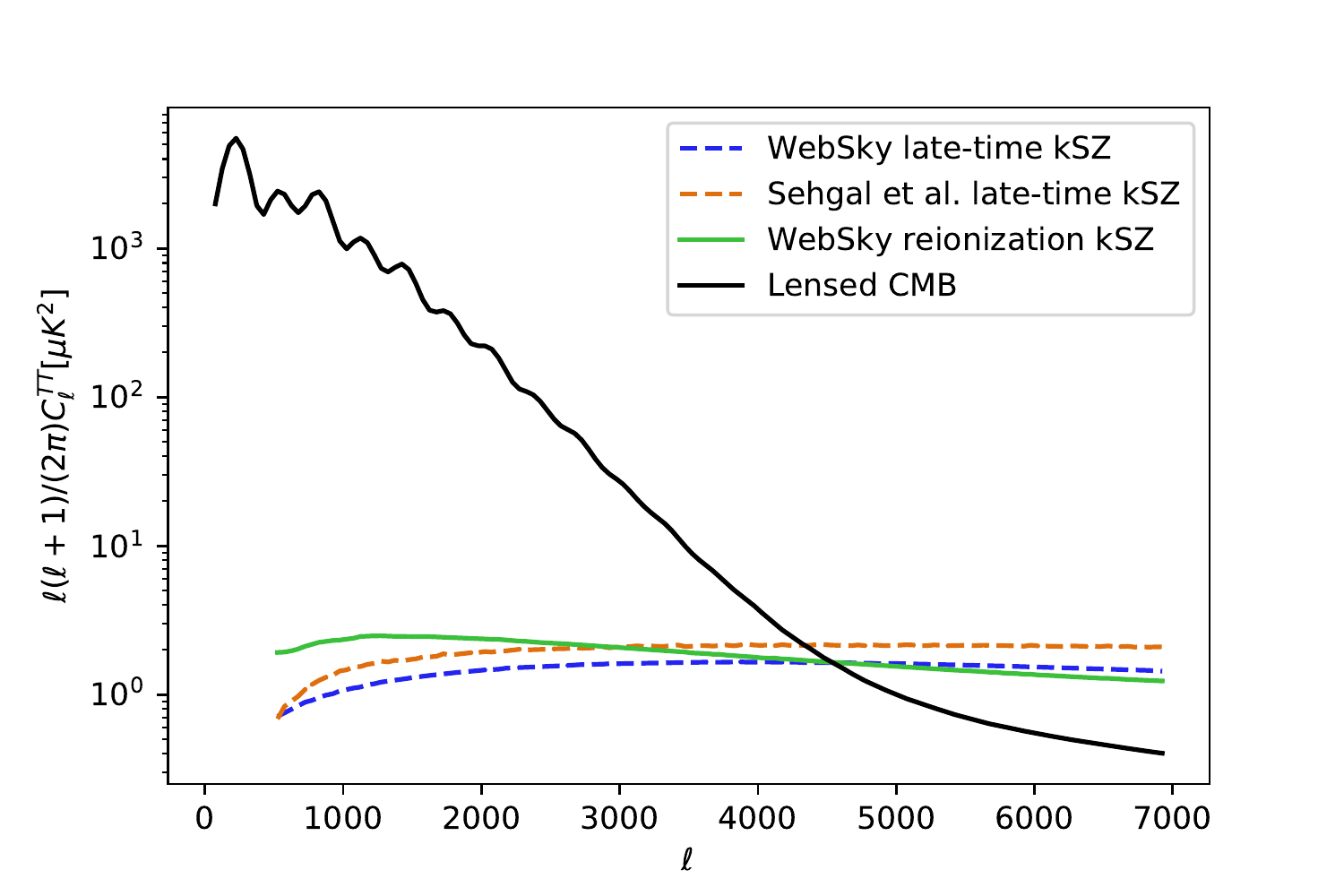}
\caption{The CMB power spectrum from lensed CMB, reionization kSZ, and late-time kSZ. The kSZ spectra are from the WebSky simulation \cite{Stein_2020} and the Sehgal et al.\ simulation \cite{Sehgal_2010}.}
\centering
\label{fig:ps}
\end{figure}

The kSZ effect is induced by the bulk motion of the ionized gas when CMB photons travel through the universe~\cite{10.1093/mnras/190.3.413}. The temperature fluctuations induced by the kSZ effect in a direction \(\hat{\mathbf{n}}\) are given by (in units with \(c=1\))
\begin{equation}
  \label{eq:ksz2}
  \frac{\Delta T^{\mathrm{kSZ}}(\hat{\mathbf{n}})}{T_{\mathrm{CMB}}} = -\sigma_{T} \int \frac{d \eta}{1+z} e^{-\tau} n_{e}(\hat{\mathbf{n}}, \eta) \mathbf{v}_{e} \cdot \hat{\mathbf{n}},
\end{equation}
where \(\sigma_{T}\) is the Thomson scattering cross section, \(n_{e}\) is the local number density of free electrons, and \(\mathbf{v}_{e} \cdot \hat{\mathbf{n}}\) is the peculiar velocity of the electrons projected along the line of sight, defined such that positive (negative) velocities point away from (toward) our vantage point. The distribution of  the kSZ temperature fluctuations is non-Gaussian due to gravitational and baryonic-feedback-induced non-linearities in the gas distribution.

The kSZ effect is dominated by epochs with large electron density fluctuations. The kSZ signal has two main contributions \cite{Park:2013mv}: the reionization contribution and the late-time contribution. The reionization kSZ arises from the local patchy and incomplete ionization during the epoch of reionization \cite{Gnedin_2001, PhysRevLett.81.2004, battaglia13_reion_large_scales}. The late-time kSZ, also known as the post-reionization kSZ \cite{10.1111/j.1365-2966.2004.07298.x}, arises after the epoch of reionization, when the universe is fully ionized.

Fig.~\ref{fig:ps} shows that on the smallest scales $\ell\gtrsim4000$, the kSZ effect starts to dominate the CMB temperature power spectrum.  The reionization and late-time kSZ power spectrum contributions are comparable in magnitude. The reionization kSZ is expected to be only weakly correlated with CMB lensing, but the correlation of the late-time kSZ and CMB lensing is much larger, for most of the lensing fluctuations are generated at relatively low redshift. However, the reionization kSZ signal is of potential concern for lensing reconstruction bias because of its intrinsically non-Gaussian pattern on the sky.

We assume secondary anisotropies have been removed by multifrequency component separation methods (or with geometric methods), and  except the kSZ and lensing signals. The integrated Sachs-Wolfe effect is also blackbody, but is negligible except on the largest angular scales.
Then the total CMB temperature map under the flat-sky approximation can be written as
\begin{equation}
  \label{eq:observed T}
  {T}^{\mathrm{tot}}(\mathbf{x}) = \tilde{T}(\mathbf{x}) + T^{\mathrm{kSZ}}(\mathbf{x}),
\end{equation}
where \(\tilde{T}(\mathbf{x})\) is the lensed CMB field and \(T^{\mathrm{kSZ}}(\mathbf{x})\) is the kSZ signal.

The total temperature map \({T}^{\mathrm{tot}}(\mathbf{x})\) contains \({T}^{\mathrm{kSZ}}(\mathbf{x})\) which is non-Gaussian and may correlate with lensing. When we perform CMB lensing reconstruction with temperature anisotropies using Eq.~\eqref{eq:kappa estimator}, we include the Fourier modes \(T^{\mathrm{kSZ}}(\bm{\ell})\) in the estimator.  The non-Gaussianity of kSZ and any kSZ-lensing correlation bias the estimation of \(C_{L}^{\kappa\kappa}\) since extra connected terms are brought to the CMB temperature trispectrum as shown in Appendix~\ref{sec:appendix A}.

Estimating this bias is straightforward. We use one set of simulated kSZ realizations, described in the next section, to represent \(T^{\mathrm{kSZ}}(\mathbf{x})\) in Eq.~\eqref{eq:observed T}, and generate another set of Gaussian kSZ realizations \(T^{\mathrm{kSZ,g}}(\mathbf{x})\) with the same average power spectrum. We also have a set of lensed CMB realizations \(\tilde{T}(\mathbf{x})\).
We define
\begin{equation}
  \label{eq:t,g}
  T^{\mathrm{tot,g}}(\mathbf{x}) = \tilde{T}(\mathbf{x}) + T^{\mathrm{kSZ,g}}(\mathbf{x})
\end{equation}
which is similar to Eq.~\eqref{eq:observed T} except for the Gaussian kSZ term. We apply the quadratic estimator in Eq.~\eqref{eq:kappa estimator} to \(T^{\mathrm{tot}}(\bm{\ell})\) and \(T^{\mathrm{tot,g}}(\bm{\ell})\) to obtain two sets of reconstructed lensing convergence maps with power spectra of \(C_{L, \mathrm{tot}}^{\hat{\kappa} \hat{\kappa}}\) and \(C_{L, \mathrm{tot,g}}^{\hat{\kappa} \hat{\kappa}}\).
Since \(T^{\mathrm{tot}}\) and \(T^{\mathrm{tot,g}}\) have the same power spectrum and share the same lensed CMB realizations, they have the same reconstruction biases shown in Eq.\eqref{eq:noise substraction} on average. Because \(T^{\mathrm{tot,g}}(\mathbf{x})\) is Gaussian and not correlated with $\tilde{T}(\mathbf{x})$, it does not induce the $(\Delta C_{L}^{\kappa \kappa})_{\mathrm{FG}}$ in Eq.~\eqref{eq:noise substraction}. Thus, the bias to CMB lensing reconstruction from temperature maps due to the kSZ effect can be estimated by
\begin{equation}
  \label{eq:ksz bias}
  (\Delta C_{L}^{\kappa \kappa})_{\mathrm{kSZ}} = \langle C_{L,\mathrm{tot}}^{\hat{\kappa} \hat{\kappa}}\rangle - \langle C_{L,\mathrm{tot,g}}^{\hat{\kappa} \hat{\kappa}} \rangle,
\end{equation}
where the brackets are the average over the two sets of reconstructed lensing convergence maps.

This method to estimate \((\Delta C_{L}^{\kappa \kappa})_{\mathrm{kSZ}}\) in Eq.~\eqref{eq:ksz bias} can be applied to both the reionization kSZ and the late-time kSZ signals \cite{ferraro18_bias_to_cmb_lensin_recon}. A similar approach has also been applied in \cite{Bohm:2018omn} to estimate the non-Gaussian lensing bias in CMB lensing reconstruction.

Since the reionization kSZ signal is only weakly correlated with the CMB lensing field\footnote{The reionization kSZ simulations used in this paper are actually uncorrelated with the CMB lensing field.}, the kSZ trispectrum induced by its non-Gaussianity is expected to be the dominant contribution to the CMB lensing reconstuction bias. Thus, we can apply the CMB lensing reconstruction algorithm to the reionization kSZ map directly without adding the lensed CMB map and compute the bias by
\begin{equation}
  \label{eq:ksz trispectrum bias}
    (\Delta C_{L}^{\kappa \kappa})_{\mathrm{kSZ}} = \langle C_{L,\mathrm{kSZ}}^{\hat{\kappa} \hat{\kappa}}\rangle - \langle C_{L,\mathrm{kSZ,g}}^{\hat{\kappa} \hat{\kappa}} \rangle.
\end{equation}
This bias estimate neglects terms arising from the correlation between the kSZ and CMB lensing fields. For the full kSZ signal including correlation of lensing with late-time kSZ,
we need to apply the CMB lensing reconstruction algorithm to the sum of the lensed CMB map and the late-time kSZ map, and estimate the bias using Eq.~\eqref{eq:ksz bias}.

\begin{figure*}[htbp!]
\includegraphics[width=\textwidth]{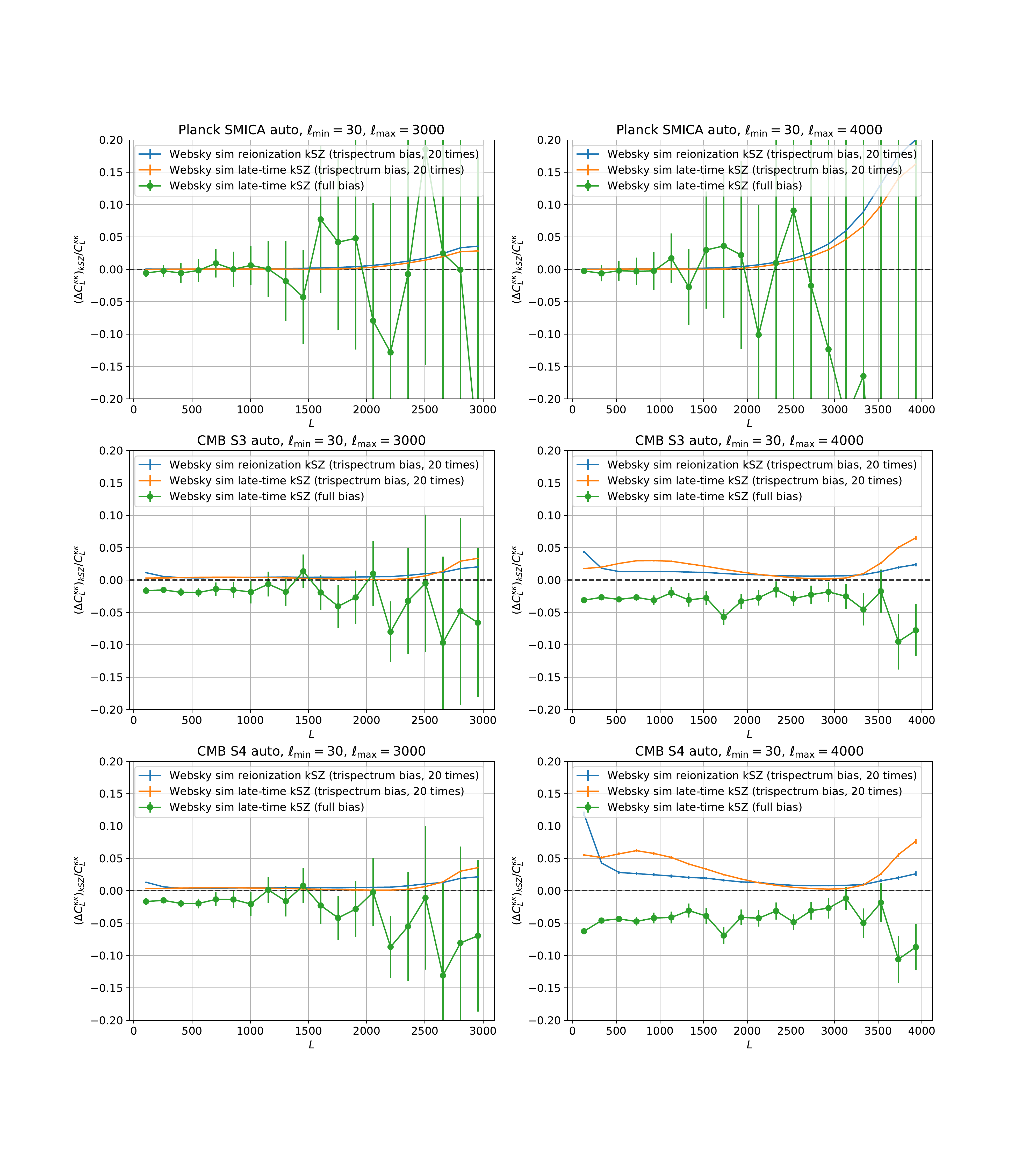}
\caption{The fractional bias to the reconstructed CMB lensing convergence power spectrum induced by the reionization kSZ trispectrum (blue curves), the late-time kSZ trispectrum (orange curves), and the full late-time kSZ bias (green curves), computed using the WebSky simulation. The curves showing the reionization kSZ trispectrum bias, the late-time kSZ trispectrum bias and the corresponding error bars are multiplied by a factor of 20 for visibility. The error bars are difficult to see by eye, even after the factor of 20 amplification.  We use the temperature multipole range for CMB lensing reconstruction from \(\ell_{\mathrm{min}} = 30\) to \(\ell_{\mathrm{max}}=3000\) (left panels) or \(\ell_{\mathrm{min}} = 30\) to \(\ell_{\mathrm{max}}=4000\) (right panels).  Experimental parameters for the Planck SMICA, CMB-S3-like, and CMB-S4-like configurations are given in Table~\ref{table:experiments}.}
  \label{fig:websky ri}
\centering
\end{figure*}

\section{Simulations and Results}
\label{sec:simulations}
As explained in Section~\ref{sec:kSZ bias}, we apply the CMB lensing reconstruction algorithm to a set of independent reionization kSZ realizations. These realizations are cut from a full-sky kSZ map provided by the WebSky\footnote{\url{https://lambda.gsfc.nasa.gov/simulation/tb_mocks_data.cfm}} extragalactic CMB simulations \cite{Stein_2020}.

The reionization kSZ temperature map is provided in blackbody thermodynamic temperature units, and is constructed from the free electron and velocity fields at \(z>5.5\). It is uncorrelated with the lensing map and the primary unlensed CMB map. The reionization kSZ simulation \cite{Alvarez:2015xzu} uses \(4096^3\) elements in a periodic box of side length 8 Gpc/$h$. Three astrophysical reionization parameters define the simulation: \(M_{\mathrm{min}}\), the minimum halo mass capable of hosting ionizing sources; \(\lambda_{\mathrm{abs}}\), the comoving absorption system (Lyman-limit absorption systems) mean free path; and \(\zeta_{\mathrm{ion}}\), the number of ionizing photons per atom escaping each halo \cite{Alvarez:2015xzu}. These parameters are chosen as \(M_{\mathrm{min}} = 10^{9}M_{\odot}\), \(\lambda_{\mathrm{abs}}=50 \, {\rm Mpc}/h \) and \(\zeta_{\mathrm{ion}}=50\), and yield a total Thomson scattering optical depth of \(\tau=0.059\) and a mean redshift of reionization of $z_{\rm re} = 7.93$ \footnote{\url{https://www.cita.utoronto.ca/~malvarez/research/ksz-data/run_params.txt}}, consistent with current constraints from \emph{Planck}~\cite{Planck:2018vyg}.

We perform the following analysis on this simulation:
\begin{enumerate}
\item
  \label{itm:step 1}
  We get one full-sky reionization kSZ map \(T^{\mathrm{kSZ}}(\hat{\bm{n}})\), where the direction vector \(\hat{\bm{n}}\) indicates a full-sky map.
\item We generate 30 full-sky Gaussian reionization kSZ maps \(T^{\mathrm{kSZ,g}}(\hat{\bm{n}})\) with the same average power spectrum as that of \(T^{\mathrm{kSZ}}(\hat{\bm{n}})\) using {\tt healpy}.\footnote{\url{https://github.com/healpy/healpy}} We use multiple Gaussian kSZ realizations to average down the statistical error.

\item
  \label{itm:step 2}
  We select regions spanning \(\pm45^{\circ}\) in declination and \(\ 360^{\circ} \) in RA from \(T^{\mathrm{kSZ}}(\hat{\bm{n}})\) and \(T^{\mathrm{kSZ,g}}(\hat{\bm{n}})\) using \texttt{pixell}\footnote{\url{https://github.com/simonsobs/pixell}}. In the pixel space, the regions correspond to 21600 pixels in width and 5400 pixels in height in our simulation. We cut 36 non-overlapping patches with 1800 pixels both in width and in height from each of the regions. So we obtain a set of 36 reionization kSZ cutouts \(T^{\mathrm{kSZ}}(\mathbf{x})\) and a set of 1080 Gaussian reionization kSZ cutouts \(T^{\mathrm{kSZ,g}}(\mathbf{x})\), where the position vector \(\mathbf{x}\) indicates we treat these cutouts under the flat-sky approximation. Because the cutouts are approximately independent regions, we consider them as independent realizations.

\item
  \label{itm:step 3}
  We run the \texttt{symlens}\footnote{\url{https://github.com/simonsobs/symlens}} flat-sky CMB lensing reconstruction algorithm on the sets of \(T^{\mathrm{kSZ}}(\mathbf{x})\) cutouts and \(T^{\mathrm{kSZ,g}}(\mathbf{x})\) cutouts, and get two sets of reconstructed convergence maps: \(\hat{\kappa}_{\mathrm{kSZ}}(\mathbf{x})\) and \(\hat{\kappa}_{\mathrm{kSZ,g}}(\mathbf{x})\). Their power spectra \(C_{L,\mathrm{kSZ}}^{\hat{\kappa} \hat{\kappa}}\) and \(C_{L,\mathrm{kSZ,g}}^{\hat{\kappa} \hat{\kappa}}\) are bandpowers binned in the two-dimensional reconstructed Fourier \(\kappa\) maps. We use a multipole bin width $\Delta{L}=150$ for the analyses with $\ell_{\mathrm{max}}=3000$ and $\Delta{L}=200$ for $\ell_{\mathrm{max}}=4000$, where $\ell_{\mathrm{max}}$ is the maximum temperature multipole for CMB lensing reconstruction.
\item
  \label{itm:step 4}
Following \cite{ferraro18_bias_to_cmb_lensin_recon}, note that we do not include the contribution of detector noise \(T^{\mathrm{det}}(\mathbf{x})\), making the results less noisy without being biased. The detector noise Fourier modes \(T^{\mathrm{det}}(\bm{\ell})\) are only included in the denominator of the weight \(g(\bm{\ell}, \mathbf{L})\) in Eq.~\eqref{eq:weight}.
\item
  \label{itm:step 5}
  The reionization kSZ-induced bias \((\Delta{C_{L}^{\kappa \kappa}})_{\mathrm{kSZ}}\) is estimated by Eq. \eqref{eq:ksz trispectrum bias}. Note that in Eq. \eqref{eq:ksz trispectrum bias}, the angle brackets in the first term indicate the average bandpower of the set of 36 \(C_{L,\mathrm{kSZ}}^{\hat{\kappa}\hat{ \kappa}}\) and the second pair is the average bandpower of the set of 1080 \(C_{L,\mathrm{kSZ,g}}^{\hat{\kappa} \hat{\kappa}}\) in our simulations.

\item \label{itm:step 6}The fractional bias is defined as \( \frac{(\Delta{{C_{L}^{\kappa \kappa}}})_{\mathrm{kSZ}}} {C_{L}^{\kappa \kappa}}\),  where \(C_{L}^{\kappa \kappa}\) is the power spectrum of the true CMB lensing convergence field. The error bar on \((\Delta{C_{L}^{\kappa \kappa}})_{\mathrm{kSZ}}\) is estimated from the Gaussian kSZ realizations.

\item As a comparison, we apply Step~\ref{itm:step 1} to Step~\ref{itm:step 4} above to the late-time kSZ simulations from WebSky to estimate the bias to CMB lensing reconstruction from the late-time kSZ trispectrum using Eq.~\eqref{eq:ksz trispectrum bias}. The late-time kSZ full-sky map from WebSky is constructed from the free electron and velocity fields at \(z<4.5\).

\item We also estimate the full bias to CMB lensing reconstruction from the late-time kSZ field, i.e., including terms from the kSZ trispectrum and the kSZ-lensing correlation, by substituting \(T^{\mathrm{ksz}}(\hat{\bm{n}})\) and \(T^{\mathrm{ksz,g}}(\hat{\bm{n}})\) with \(T^{\mathrm{tot}}(\hat{\bm{n}})\) and \(T^{\mathrm{tot,g}}(\hat{\bm{n}})\) in Step~\ref{itm:step 1}, where \(T^{\mathrm{tot}}(\hat{\bm{n}}) = \tilde{T}(\hat{\bm{n}}) + T^{\mathrm{kSZ}}(\hat{\bm{n}})\) and \(T^{\mathrm{tot,g}}(\hat{\bm{n}}) =  \tilde{T}(\hat{\bm{n}}) + T^{\mathrm{kSZ,g}}(\hat{\bm{n}})\). \(\tilde{T}(\hat{\bm{n}})\) is a full-sky lensed CMB temperature map and $T^{\mathrm{kSZ}}(\hat{\bm{n}})$ refers to the late-time kSZ full-sky map from WebSky. The full bias from the late-time kSZ field is estimated using Eq.~\eqref{eq:ksz bias}.

\end{enumerate}

In Fig.~\ref{fig:websky ri}, we show the fractional bias to the CMB lensing power spectrum induced by the reionization kSZ trispectrum, the late-time kSZ trispectrum, and the full late-time kSZ for Planck, CMB-S3-like, and CMB-S4-like experiments with $\ell_{\mathrm{max}} = 3000$ or $\ell_{\mathrm{max}} = 4000$, where ${\ell}_{\mathrm{max}}$ is the maximum temperature multipole used in the CMB lensing reconstruction.  The experimental configurations are defined in Table \ref{table:experiments}; in order to facilitate comparison, we adopt the same settings as used in~\cite{ferraro18_bias_to_cmb_lensin_recon}.  The curves showing the bias from the reionization kSZ trispectrum and the late-time kSZ trispectrum in Fig.~\ref{fig:websky ri} have been multiplied by a factor of 20 for visibility.  We can see the fractional bias from the reionization kSZ trispectrum is positive and smaller than $0.25\%$ at most scales, except for the largest scales when considering $\ell_{\mathrm{max}}=4000$. The late-time kSZ trispectrum produces a comparable bias to that from the reionization kSZ trispectrum. The full bias from the late-time kSZ signal is at least one order of magnitude larger than the other two. For all three cases, the bias with $\ell_{\mathrm{max}}=4000$ is larger than that with $\ell_{\mathrm{max}}=3000$, as expected since more foreground-contaminated modes are used in the former case. Since the reionization kSZ field is only weakly correlated with the CMB lensing field, it is safe to use the bias from the reionization kSZ trispectrum to approximate the full bias from the reionization kSZ field.  Thus the full reionization kSZ bias is much less significant than the full late-time kSZ bias to the reconstructed CMB lensing power spectrum.

\begin{table}[t!]
\centering
\begin{tabular}{|p{2.4cm}|p{2.4cm}|p{2.4cm}|}
 \hline
 CMB Experiment & Noise Level \({\Delta}_{T}\ [\mathrm{\mu{\rm K} \, arcmin}]\) & Beam FWHM \({\theta}_{\mathrm{FWHM}} \ [\mathrm{arcmin}]\)\\[0.5ex]
  \hline
  \hline
  Planck SMICA& 45 & 5 \\
  \hline
  CMB-S3-like & 7 & 1.4 \\
  \hline
  CMB-S4-like & 1 & 3 \\
 \hline
\end{tabular}
\caption{Experimental configurations. Note that the actual beam FWHM for the CMB-S4 reference design is 1.4 arcmin \cite{Abazajian:2019eic}. We use 3 arcmin in this work to facilitate comparison with \cite{ferraro18_bias_to_cmb_lensin_recon}.}
 \label{table:experiments}
\end{table}

\begin{figure}[t!]
\includegraphics[width=1.0\linewidth]{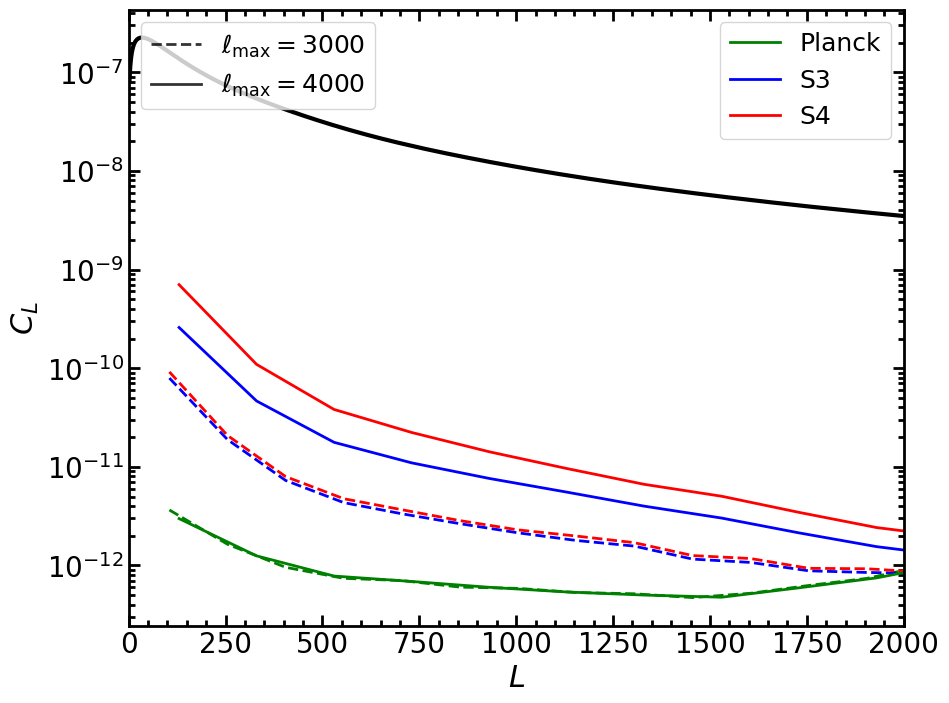}
\caption{The bias to the reconstructed CMB lensing convergence power spectrum from reionization kSZ trispectrum using \(\ell_{\mathrm{max}}=3000\) (dashed lines) and \(\ell_{\mathrm{max}}=4000\) (solid lines) for Planck SMICA, CMB-S3-like, and CMB-S4-like experiments. The true power spectrum of the CMB lensing convergence is also shown (black solid line).}
\centering
\label{fig:ksz trispectrum bias}
\end{figure}

In Fig.~\ref{fig:ksz trispectrum bias}, we show the absolute bias from the reionization kSZ trispectrum to the reconstructed CMB lensing convergence power spectrum for different experiments along with the true power spectrum of CMB lensing convergence. Within this $L$ range, the reionization kSZ-induced bias is at least two orders of magnitude lower than the lensing convergence power spectrum for CMB-S3-like and CMB-S4-like experiments using $\ell_{\mathrm{max}}=4000$, three orders of magnitude lower for CMB-S3 and CMB-S4 with $\ell_{\mathrm{max}}=3000$, and four orders of magnitude lower for Planck SMICA.

\begin{figure*}[htbp]
\includegraphics[width=\textwidth]{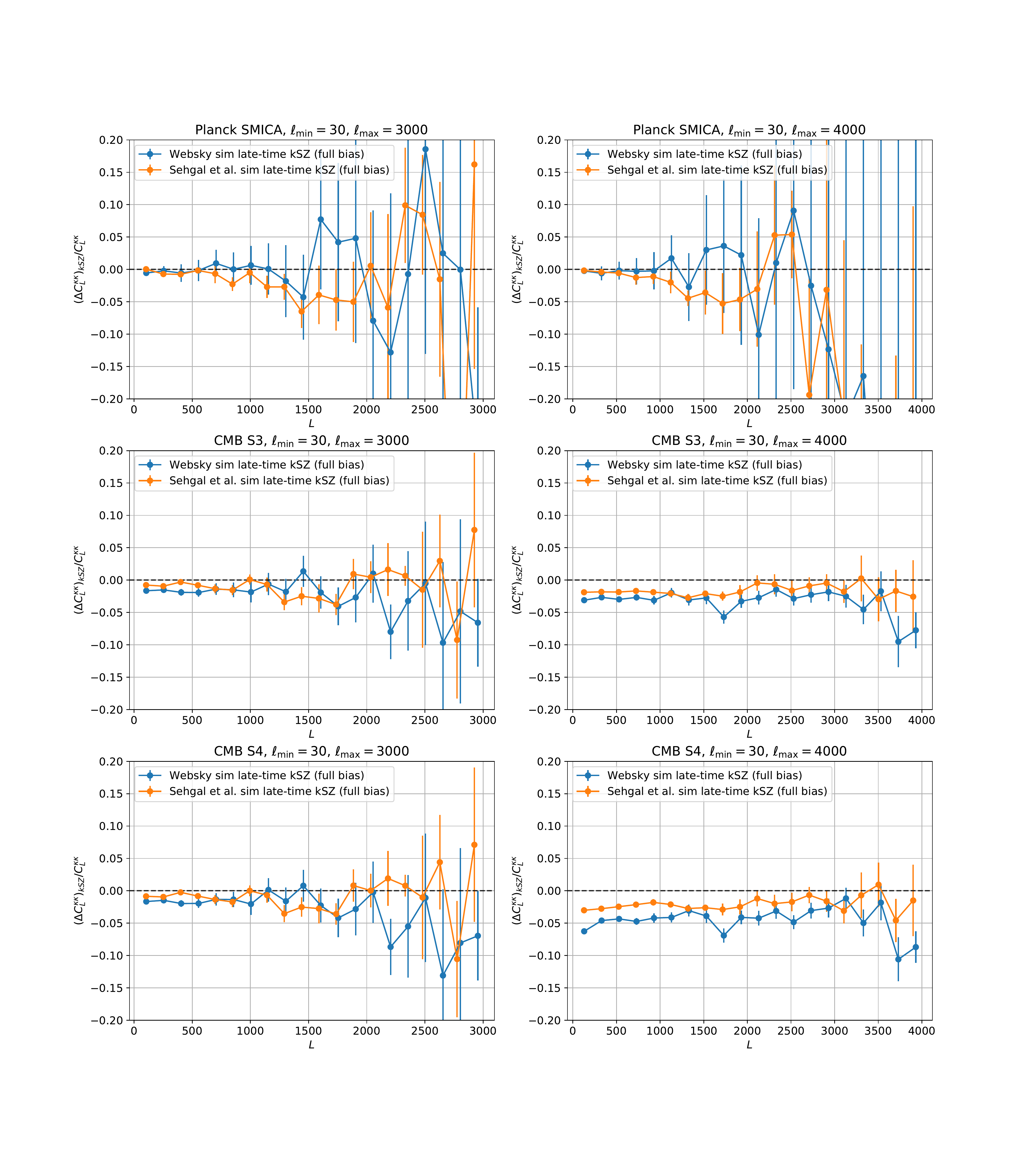}
  \caption{The full late-time-kSZ-induced fractional bias to the reconstructed CMB lensing convergence power spectrum computed using the WebSky simulation~\cite{Stein_2020} (blue curves) and using the Sehgal et al.~simulation~\cite{Sehgal_2010} (orange curves, taken from~\cite{ferraro18_bias_to_cmb_lensin_recon}). The experimental settings and temperature multipole ranges used in the lensing reconstruction are identical to those in Fig.~\ref{fig:websky ri}.  We find that the WebSky-predicted biases are up to twice as large as those predicted by the Sehgal et al.~simulation.} \label{fig:late time full}
  \centering
\end{figure*}

\begin{figure}[t!]
\includegraphics[width=1.0\linewidth]{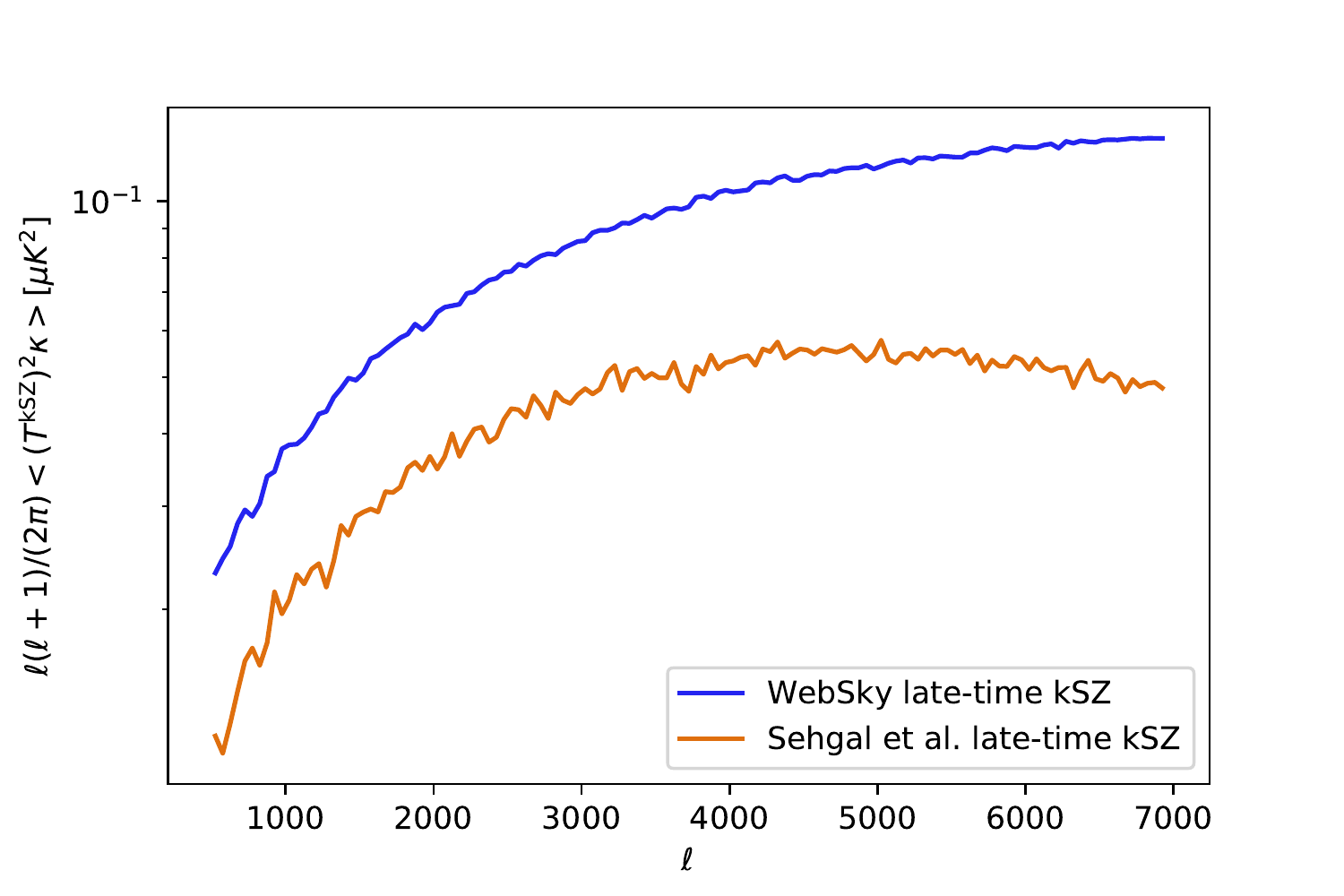}
\caption{A projection of the $\left\langle T^{\mathrm{kSZ}}T^{\mathrm{kSZ}}\kappa\right\rangle$ bispectrum estimated from the cross-correlation of  $T_{\mathrm{kSZ}}^2$ and $\kappa$, shown for each of the WebSky late-time kSZ (blue) and the Sehgal et al. late-time kSZ (orange) simulations. The curves have been binned with $\Delta \ell = 50$. Despite lower two-point power in the kSZ anisotropies for the WebSky simulation, the cross-bispectrum with CMB lensing is larger, which agrees with our observation that the bias to CMB lensing is larger in the WebSky simulation.}
\centering
\label{fig:cross ps}
\end{figure}

In Fig.~\ref{fig:late time full}, we show the full kSZ-induced fractional bias to the CMB lensing power spectrum computed using the WebSky simulation and using the Sehgal et al. simulation\cite{Sehgal_2010}.  The latter result is taken directly from \cite{ferraro18_bias_to_cmb_lensin_recon}. The CMB multipole ranges and experimental configurations are the same as those in Fig.~\ref{fig:ksz trispectrum bias}. For both sets of results, the full late-time kSZ-induced bias is smaller than the statistical error bars for Planck. For CMB-S3 and CMB-S4, the bias is negative and about several percent for $\ell_{\mathrm{max}}=4000$, and about half that for $\ell_{\mathrm{max}}=3000$. For CMB-S3 and CMB-S4 with $\ell_{\mathrm{max}}=4000$, the bias from the WebSky simulation is about 1.5 to 2 times of that computed using the Sehgal et al. simulation. This is consistent with the result in Fig.~\ref{fig:cross ps}, which shows that the bispectrum of  $<T^{\mathrm{kSZ}}T^{\mathrm{kSZ}}\kappa>$ is larger in the WebSky simulation than in the Sehgal et al. simulation, where $T^{\mathrm{kSZ}}$ refers to the late-time kSZ signal and $\kappa$ is the CMB lensing convergence. This bispectrum appears in Eq.~\eqref{eq:bispectrum contraction} in Appendix~\ref{sec:appendix A}, which is the largest overall connected term of CMB trispectrum contributing to the late-time kSZ-induced bias. The difference in these predicted biases also reflects the current uncertainty in our understanding of the late-time kSZ field, and indicates that data-driven methods should be used to mitigate the kSZ bias, rather than methods assuming particular theoretical models.

\section{Discussion and Conclusion}
\label{sec:discussion and conclusions}
The kSZ effect is the largest blackbody contaminant to the CMB primary temperature anisotropies, and it cannot be removed by multifrequency component separation techniques. The kSZ signal has two contributions, the late-time contribution and the reionization contribution. In this paper, we focus on the bias to the reconstructed CMB lensing convergence power spectrum induced by the reionization kSZ signal. Since the reionization kSZ field is only weakly correlated with the CMB lensing field, its trispectrum should be the dominant contribution to the bias to CMB lensing reconstruction. We estimate this bias by applying the flat-sky CMB lensing reconstruction algorithm to reionization kSZ simulations and corresponding Gaussian realizations with the same power spectrum. We also apply the same method to estimate the bias to CMB lensing reconstruction from the late-time kSZ trispectrum alone, and the full late-time kSZ-induced bias.

Using the WebSky simulation, we find that the fractional bias from the reionization kSZ signal is positive and smaller than $0.25\%$ at $L<3000$ when using $\ell_{\mathrm{max}}=4000$ for CMB-S3-like and CMB-S4-like experiments, and can be even smaller when using $\ell_{\mathrm{max}}=3000$. The fractional bias computed using the WebSky late-time kSZ field is more than 10 times larger than that from the reionization kSZ field, which implies that the latter one is negligible for ongoing and upcoming experiments. These conclusions are based on the current numerical simulations of reionization from WebSky, and results may differ somewhat for other reionization models; however, the reionization-induced bias is very unlikely to be comparable to that from the late-time kSZ field for any reasonable reionization model.

As a comparison, we show that the bias induced by the late-time kSZ trispectrum is comparable to that induced by the reionization kSZ field, which are both much smaller than the full late-time kSZ bias. Thus, for the late-time kSZ, the kSZ-lensing correlation contributes much more than the kSZ trispectrum, as also found earlier in~\cite{ferraro18_bias_to_cmb_lensin_recon} (see their Fig.~6).

In addition, we compare the bias from the late-time kSZ field computed using the WebSky simulation to that computed using the Sehgal et al. simulation in \cite{ferraro18_bias_to_cmb_lensin_recon}. For CMB-S3 and CMB-S4 with $\ell_{\mathrm{max}}=4000$, we find that the absolute value of the former is about 1.5 to 2 times larger than the latter. Considering the statistical precision of CMB-S3 and CMB-S4, we confirm that the bias to CMB lensing reconstruction from the late-time kSZ effect is non-negligible and requires mitigation techniques, such as foreground-hardened estimators \cite{Namikawa_2013, Osborne:2013nna, Sailer:2020lal} or shear-only reconstruction \cite{Schaan:2018tup}. The former technique effectively deprojects a point-source-like trispectrum from the lensing power spectrum measurement, which is expected to work well for the late-time kSZ signal. For the latter technique, since the lensing shear (local quadrupolar distortion) is less degenerate with the extragalactic foregrounds, the shear-only reconstruction is less sensitive to foregrounds. Applying the shear-only technique to mitigate the kSZ-induced bias was done for the Sehgal et al.~kSZ bias in~\cite{Schaan:2018tup}; we leave similar analysis for the WebSky simulations to future work.

\section*{Acknowledgments}
We thank Marcelo Alvarez and Yilun Guan for useful discussions and comments. This work uses maps of The WebSky Extragalactic CMB Simulations \cite{Stein_2020}, resources of the National Energy Research Scientific Computing Center, the \texttt{symlens} flat-sky CMB lensing reconstruction algorithm,\footnote{\url{https://github.com/simonsobs/symlens}} the \texttt{healpy} package,\footnote{\url{https://github.com/healpy/healpy}} and the \texttt{pixell} package.\footnote{\url{https://github.com/simonsobs/pixell}} Research at Perimeter Institute is supported in part by the Government of Canada through the Department of Innovation, Science and Industry Canada and by the Province of Ontario through the Ministry of Colleges and Universities.  JCH acknowledges support from NSF grant AST-2108536.

\appendix
\section{kSZ bias to CMB temparature trispectrum}
\label{sec:appendix A}
The power spectrum of $ \hat\kappa$ can be written as
\begin{equation}
  \label{eq:kappa ps}
  \begin{aligned}
  \langle \hat\kappa({\mathbf{L}})\hat\kappa({\mathbf{L^{\prime}}}) \rangle &=  \frac{1}{4} L^{2} L'^{2} A(\mathbf{L})A(\mathbf{L'})\\
  &\int_{\bm{\ell}_{1}, \bm{\ell}_{2},\bm{\ell}_{3}, \bm{\ell}_{4}}g(\bm{\ell}_{1},\bm{\ell}_{1}+\bm{\ell}_{2})g(\bm{\ell}_{3},\bm{\ell}_{3}+\bm{\ell}_{4})\\
  &\langle T^{\mathrm{tot}}(\bm{\ell}_{1})T^{\mathrm{tot}}(\bm{\ell}_{2})T^{\mathrm{tot}}(\bm{\ell}_{3})T^{\mathrm{tot}}(\bm{\ell}_{4})\rangle\\
  &\delta(\mathbf{L}-\bm{\ell}_{1}-\bm{\ell}_{2}) \delta(\mathbf{L}-\bm{\ell}_{3}-\bm{\ell}_{4})
\end{aligned}
\end{equation}
To understand the bias to reconstructed CMB lensing power spectrum from the late-time kSZ and the reionization kSZ, we check the trispectrum
\begin{equation}
  \label{eq:trispectrum}
 \langle T^{\mathrm{tot}}(\bm{\ell}_{1})T^{\mathrm{tot}}(\bm{\ell}_{2})T^{\mathrm{tot}}(\bm{\ell}_{3})T^{\mathrm{tot}}(\bm{\ell}_{4}) \rangle,
\end{equation}
where we decompose the \(T^{\mathrm{tot}}(\bm{\ell})\) as
\begin{equation}
  \label{eq:observed T fourier}
  {T}^{\mathrm{tot}}(\bm{\ell}) = \tilde{T}(\bm{\ell}) + T^{\mathrm{kSZ}}(\bm{\ell}).
\end{equation}
Again, note that we do not include \(T^{\mathrm{det}}\), since it is Gaussian with zero expectation and uncorrelated with any other components.

The lensed CMB temperature Fourier modes can be expressed as
\begin{equation}
  \label{eq:lensed fourier}
  \tilde{T}(\bm{\ell}) = T(\bm{\ell}) + \delta T(\bm{\ell}) + \delta^{2}T(\bm{\ell}) + \mathcal{O}(\phi^{3}(\bm{\ell}))
\end{equation}
with \(\mathcal{O}(\phi)\) correction
\begin{equation}
  \label{eq:1st order correction}
\delta T(\bm{\ell}) = -\int_{\bm{\ell}^{\prime}} \bm{\ell}^{\prime}\cdot(\bm{\ell}-\bm{\ell}^{\prime})T(\bm{\ell}^{\prime})\phi(\bm{\ell}-\bm{\ell}^{\prime}),
\end{equation}
and \(\mathcal{O}(\phi^{2})\) correction
\begin{equation}
    \label{eq:2nd order correction}
  \begin{aligned}
  \delta^{2} T(\bm{\ell})=\frac{1}{2} \int_{\bm{\ell}^{\prime}} \int_{\bm{\ell}^{\prime \prime}}\left[\bm{\ell}^{\prime} \cdot \bm{\ell}^{\prime \prime}\right]&\left[\bm{\ell}^{\prime} \cdot\left(\bm{\ell}-\bm{\ell}^{\prime}-\bm{\ell}^{\prime \prime}\right)\right] T\left(\bm{\ell}^{\prime}\right) \\
  &\phi\left(\bm{\ell}^{\prime \prime}\right) \phi\left(\bm{\ell}-\bm{\ell}^{\prime}-\bm{\ell}^{\prime \prime}\right),
  \end{aligned}
\end{equation}
where $\phi$ is the CMB lensing potential.

In our calculation, we only include \(\delta T(\bm{\ell})\). We assume that the unlensed CMB is a Gaussian field. \(\phi\) and \(T^{\mathrm{kSZ}}\) are both not Gaussian. We ignore the ISW effect and the Rees-Sciama effect, so we do not consider the correlation of unlensed CMB with lensing \(C^{T\phi}\).

We plug Eq.~\eqref{eq:t,g}, Eq.~\eqref{eq:1st order correction} and Eq.~\eqref{eq:2nd order correction} into Eq.~\eqref{eq:trispectrum}, and check all the possible contractions. Any terms including odd powers of \(T^{\mathrm{kSZ}}\) vanish on average according to its symmetry~\cite{ferraro18_bias_to_cmb_lensin_recon}. The fields labeled with primes correspond to the second reconstruction field \(\hat{\kappa}(\mathbf{L'})\) in Eq.~\eqref{eq:kappa ps}, and the fields without primes correspond to the first one.

There are several types of contraction up to the order of \(\phi^{2}\) with even power of \(T^{\mathrm{kSZ}}\):

type a:
\begin{equation}
  \label{eq:bispectrum contraction}
  \begin{aligned}
    &\langle T_{\bm{\ell}_{1}}^{\mathrm{kSZ}} T_{\bm{\ell}_{2}}^{\mathrm{kSZ}} T_{\bm{\ell}_{3}}^{\prime}\phi_{\bm{\ell}_{3}}^{\prime} T_{\bm{\ell}_{4}}^{\prime} \rangle \\
    =&
  {\contraction{\langle T_{\bm{\ell}_{1}}^{\mathrm{kSZ}} T_{\bm{\ell}_{2}}^{\mathrm{kSZ}}}{T}{_{\bm{\ell}_{3}}\phi_{\bm{\ell}_{3}}^{\prime}}{T}
    \bcontraction{\langle}{T}{^{\mathrm{kSZ}}}{T}
    \bcontraction{\langle T_{\bm{\ell}_{1}}^{\mathrm{kSZ}}}{T}{_{\bm{\ell}_{2}}^{\mathrm{kSZ}}T_{\bm{\ell}_{3}}^{\prime}}{\phi}
    \langle T_{\bm{\ell}_{1}}^{\mathrm{kSZ}} T_{\bm{\ell}_{2}}^{\mathrm{kSZ}} T_{\bm{\ell}_{3}}^{\prime}\phi_{\bm{\ell}_{3}}^{\prime} T_{\bm{\ell}_{4}}^{\prime} \rangle}\\
  =&\langle T^{\prime}_{\bm{\ell}_{3}}T^{\prime}_{\bm{\ell}_{4}} \rangle {\langle T_{\bm{\ell}_{1}}^{\mathrm{kSZ}} T_{\bm{\ell}_{2}}^{\mathrm{kSZ}} \phi_{\bm{\ell}_{3}}^{\prime}\rangle} \rightarrow \mathrm{(a1)}\\
  \end{aligned}
\end{equation}

type b:
\begin{equation}
  \label{eq:ksz-phi contraction 1}
  \begin{aligned}
    &\langle T_{\bm{\ell}_{1}}^{\mathrm{kSZ}} T_{\bm{\ell}_{2}}^{\mathrm{kSZ}} T_{\bm{\ell}_{3}}^{\prime}\phi_{\bm{\ell}_{3}}^{\prime} T_{\bm{\ell}_{4}}^{\prime}\phi_{\bm{\ell}_{4}}^{\prime} \rangle \\
    =&
  {\contraction{\langle T_{\bm{\ell}_{1}}^{\mathrm{kSZ}} T_{\bm{\ell}_{2}}^{\mathrm{kSZ}}}{T}{_{\bm{\ell}_{3}}\phi_{\bm{\ell}_{3}}^{\prime}}{T}
    \bcontraction{\langle}{T}{^{\mathrm{kSZ}}}{T}
    \bcontraction{\langle T_{\bm{\ell}_{1}}^{\mathrm{kSZ}}}{T}{_{\bm{\ell}_{2}}^{\mathrm{kSZ}}T_{\bm{\ell}_{3}}^{\prime}}{\phi}
    \bcontraction{\langle T_{\bm{\ell}_{1}}^{\mathrm{kSZ}}T_{\bm{\ell}_{2}}^{\mathrm{kSZ}} T_{\bm{\ell}_{3}}^{\prime}}{\phi}{_{\bm{\ell}_{3}}^{\prime}T_{\bm{\ell}_{4}}^{\prime}}{\phi}
    \langle T_{\bm{\ell}_{1}}^{\mathrm{kSZ}} T_{\bm{\ell}_{2}}^{\mathrm{kSZ}} T_{\bm{\ell}_{3}}^{\prime}\phi_{\bm{\ell}_{3}}^{\prime} T_{\bm{\ell}_{4}}^{\prime}\phi_{\bm{\ell}_{4}}^{\prime} \rangle}\\
  =&
  \contraction{\langle T_{\bm{\ell}_{1}}^{\mathrm{kSZ}} T_{\bm{\ell}_{2}}^{\mathrm{kSZ}}}{T}{_{\bm{\ell}_{3}}^{\prime}\phi_{\bm{\ell}_{3}}^{\prime}}{T}
    \bcontraction{\langle}{T}{_{\bm{\ell}_{1}}^{\mathrm{kSZ}}}{T}
    \bcontraction{\langle T_{\bm{\ell}_{1}}^{\mathrm{kSZ}}T_{\bm{\ell}_{2}}^{\mathrm{kSZ}} T_{\bm{\ell}_{3}}^{\prime}}{\phi}{_{\bm{\ell}_{3}}^{\prime}T_{\bm{\ell}_{4}}^{\prime}}{\phi}
    \langle {T_{\bm{\ell}_{1}}^{\mathrm{kSZ}} T_{\bm{\ell}_{2}}^{\mathrm{kSZ}} T_{\bm{\ell}_{3}}^{\prime}\phi_{\bm{\ell}_{3}}^{\prime} T_{\bm{\ell}_{4}}^{\prime}\phi_{\bm{\ell}_{4}}^{\prime} \rangle} \rightarrow{\mathrm{(b1)}}\\
  +&
  \contraction{\langle T_{\bm{\ell}_{1}}^{\mathrm{kSZ}} T_{\bm{\ell}_{2}}^{\mathrm{kSZ}}}{T}{_{\bm{\ell}_{3}}^{\prime}\phi_{\bm{\ell}_{3}}^{\prime}}{T}
    \bcontraction{\langle}{T}{_{\bm{\ell}_{1}}^{\mathrm{kSZ}}T_{\bm{\ell}_{2}}^{\mathrm{kSZ}}T_{\bm{\ell}_{3}}^{\prime}}{\phi}
    \bcontraction[2ex]{\langle T_{\bm{\ell}_{1}}^{\mathrm{kSZ}}} {T}{_{\bm{\ell}_{2}}^{\mathrm{kSZ}}T_{\bm{\ell}_{3}}^{\prime}\phi_{\bm{\ell}_{3}}^{\prime}T_{\bm{\ell}_{4}}^{\prime}}{\phi}
    \langle {T_{\bm{\ell}_{1}}^{\mathrm{kSZ}} T_{\bm{\ell}_{2}}^{\mathrm{kSZ}} T_{\bm{\ell}_{3}}^{\prime}\phi_{\bm{\ell}_{3}}^{\prime} T_{\bm{\ell}_{4}}^{\prime}\phi_{\bm{\ell}_{4}}^{\prime} \rangle} \rightarrow{\mathrm{(b2)}}\\
  +&
  \contraction{\langle T_{\bm{\ell}_{1}}^{\mathrm{kSZ}} T_{\bm{\ell}_{2}}^{\mathrm{kSZ}}}{T}{_{\bm{\ell}_{3}}^{\prime}\phi_{\bm{\ell}_{3}}^{\prime}}{T}
  \bcontraction{\langle}{T}{_{\bm{\ell}_{1}}^{\mathrm{kSZ}}T_{\bm{\ell}_{2}}^{\mathrm{kSZ}}T_{\bm{\ell}_{3}}^{\prime}\phi_{\bm{\ell}_{3}}^{\prime}T_{\bm{\ell}_{4}}^{\prime}}{\phi}
 \bcontraction[2ex]{\langle T_{\bm{\ell}_{1}}^{\mathrm{kSZ}}} {T}{_{\bm{\ell}_{2}}^{\mathrm{kSZ}}T_{\bm{\ell}_{3}}^{\prime}}{\phi}
    \langle {T_{\bm{\ell}_{1}}^{\mathrm{kSZ}} T_{\bm{\ell}_{2}}^{\mathrm{kSZ}} T_{\bm{\ell}_{3}}^{\prime}\phi_{\bm{\ell}_{3}}^{\prime} T_{\bm{\ell}_{4}}^{\prime}\phi_{\bm{\ell}_{4}}^{\prime} \rangle} \rightarrow \mathrm{(b3)}\\
  +&\langle T^{\prime}_{\bm{\ell}_{3}}T^{\prime}_{\bm{\ell}_{4}} \rangle {\langle T_{\bm{\ell}_{1}}^{\mathrm{kSZ}} T_{\bm{\ell}_{2}}^{\mathrm{kSZ}} \phi_{\bm{\ell}_{3}}^{\prime} \phi_{\bm{\ell}_{4}}^{\prime} \rangle}_{\mathrm{c}} \rightarrow \mathrm{(b4)}
  \end{aligned}
\end{equation}

type c:
\begin{equation}
  \label{eq:ksz-phi contraction 2}
  \begin{aligned}
    &\langle T_{\bm{\ell}_{1}}^{\mathrm{kSZ}} T_{\bm{\ell}_{2}} \phi_{\bm{\ell}_{2}} T^{\prime \mathrm{kSZ}}_{\bm{\ell}_{3}} T_{\bm{\ell}_{4}}^{\prime}\phi_{\bm{\ell}_{4}}^{\prime} \rangle\\ =&{\contraction{\langle T_{\bm{\ell}_{1}}^{\mathrm{kSZ}} }{T}{_{\bm{\ell}_{2}}\phi_{\bm{\ell}_{2}} T_{\bm{\ell}_{3}}^{\prime \mathrm{kSZ}}}{T}
      \bcontraction{\langle}{T}{_{\bm{\ell}_{1}}^{\mathrm{kSZ}}T_{\bm{\ell}_{2}}}{\phi}
      \bcontraction{\langle T^{\mathrm{kSZ}}T_{\bm{\ell}_{1}}}{\phi}{_{\bm{\ell}_{2}}}{T}
      \bcontraction{\langle T_{\bm{\ell}_{1}}^{\mathrm{kSZ}} T_{\bm{\ell}_{2}} \phi_{\bm{\ell}_{2}}}{T}{_{\bm{\ell}_{3}}^{\prime \mathrm{kSZ}} T_{\bm{\ell}_{4}}^{\prime}}{\phi}
      \langle T_{\bm{\ell}_{1}}^{\mathrm{kSZ}} T_{\bm{\ell}_{2}} \phi_{\bm{\ell}_{2}} T_{\bm{\ell}_{3}}^{\prime \mathrm{kSZ}} T_{\bm{\ell}_{4}}^{\prime}\phi_{\bm{\ell}_{4}}^{\prime} \rangle} \\
    =&{\contraction{\langle T_{\bm{\ell}_{1}}^{\mathrm{kSZ}} }{T}{_{\bm{\ell}_{2}}\phi_{\bm{\ell}_{2}} T_{\bm{\ell}_{3}}^{\prime \mathrm{kSZ}}}{T}
      \bcontraction{\langle}{T}{_{\bm{\ell}_{1}}^{\mathrm{kSZ}}T_{\bm{\ell}_{2}}\phi_{\bm{\ell}_{2}}}{T}
      \bcontraction[2ex]{\langle T_{\bm{\ell}_{1}}^{\mathrm{kSZ}} T_{\bm{\ell}_{2}}}{\phi}{_{\bm{\ell}_{2}}T_{\bm{\ell}_{3}}^{\prime \mathrm{kSZ}} T_{\bm{\ell}_{4}}^{\prime}}{\phi}
      \langle T_{\bm{\ell}_{1}}^{\mathrm{kSZ}} T_{\bm{\ell}_{2}} \phi_{\bm{\ell}_{2}} T_{\bm{\ell}_{3}}^{\prime \mathrm{kSZ}} T_{\bm{\ell}_{4}}^{\prime}\phi_{\bm{\ell}_{4}}^{\prime} \rangle}\rightarrow \mathrm{(c1)}\\
    +&{\contraction{\langle T_{\bm{\ell}_{1}}^{\mathrm{kSZ}} }{T}{_{\bm{\ell}_{2}}\phi_{\bm{\ell}_{2}} T_{\bm{\ell}_{3}}^{\prime \mathrm{kSZ}}}{T}
      \bcontraction{\langle}{T}{_{\bm{\ell}_{1}}^{\mathrm{kSZ}}T_{\bm{\ell}_{2}}}{\phi}
      \bcontraction{\langle T_{\bm{\ell}_{1}}^{\mathrm{kSZ}} T_{\bm{\ell}_{2}} \phi_{\bm{\ell}_{2}}}{T}{_{\bm{\ell}_{3}}^{\prime \mathrm{kSZ}} T_{\bm{\ell}_{4}}^{\prime}}{\phi}
      \langle T_{\bm{\ell}_{1}}^{\mathrm{kSZ}} T_{\bm{\ell}_{2}} \phi_{\bm{\ell}_{2}} T_{\bm{\ell}_{3}}^{\prime \mathrm{kSZ}} T_{\bm{\ell}_{4}}^{\prime}\phi_{\bm{\ell}_{4}}^{\prime} \rangle}\rightarrow \mathrm{(c2)}\\
    +&{\contraction{\langle T_{\bm{\ell}_{1}}^{\mathrm{kSZ}} }{T}{_{\bm{\ell}_{2}}\phi_{\bm{\ell}_{2}} T_{\bm{\ell}_{3}}^{\prime \mathrm{kSZ}}}{T}
      \bcontraction{\langle}{T}{_{\bm{\ell}_{1}}^{\mathrm{kSZ}}T_{\bm{\ell}_{2}} \phi_{\bm{\ell}_{2}} T_{\bm{\ell}_{3}}^{\prime \mathrm{kSZ}} T_{\bm{\ell}_{4}}^{\prime}}{\phi}
      \bcontraction[2ex]{\langle T_{\bm{\ell}_{1}}^{\mathrm{kSZ}} T_{\bm{\ell}_{2}} }{\phi}{_{\bm{\ell}_{2}}}{T}
      \langle T_{\bm{\ell}_{1}}^{\mathrm{kSZ}} T_{\bm{\ell}_{2}} \phi_{\bm{\ell}_{2}} T_{\bm{\ell}_{3}}^{\prime \mathrm{kSZ}} T_{\bm{\ell}_{4}}^{\prime}\phi_{\bm{\ell}_{4}}^{\prime} \rangle}\rightarrow \mathrm{(c3)}\\
    +&\langle T_{\bm{\ell}_{2}}T_{\bm{\ell}_{4}}^{\prime} \rangle {\langle T_{\bm{\ell}_{1}}^{\mathrm{kSZ}} \phi_{\bm{\ell}_{2}} T_{\bm{\ell}_{3}}^{\prime \mathrm{kSZ}} \phi_{\bm{\ell}_{4}}^{\prime} \rangle}_{\mathrm{c}} \rightarrow \mathrm{(c4)}
  \end{aligned}
\end{equation}

type d:
\begin{equation}
  \label{eq:ksz trispectrum}
  \begin{aligned}
    &\langle T_{\bm{\ell}_{1}}^{\mathrm{kSZ}}T_{\bm{\ell}_{2}}^{\mathrm{kSZ}}T_{\bm{\ell}_{3}}^{\prime \mathrm{kSZ}}T_{\bm{\ell}_{4}}^{\prime \mathrm{kSZ}} \rangle\\
    =&\ {\contraction{\langle}{T}{_{\bm{\ell}_{1}}^{\mathrm{kSZ}}}{T}
      \contraction{\langle T_{\bm{\ell}_{1}}^{\mathrm{kSZ}}T_{\bm{\ell}_{2}}^{\mathrm{kSZ}}}{T}{_{\bm{\ell}_{3}}^{\prime \mathrm{kSZ}}}{T}
      \langle T_{\bm{\ell}_{1}}^{\mathrm{kSZ}}T_{\bm{\ell}_{2}}^{\mathrm{kSZ}}T_{\bm{\ell}_{3}}^{\prime \mathrm{kSZ}}T_{\bm{\ell}_{4}}^{\prime \mathrm{kSZ}} \rangle} \rightarrow \mathrm{(d1)}\\
    +&\ {\contraction{\langle}{T}{_{\bm{\ell}_{1}}^{\mathrm{kSZ}}T_{\bm{\ell}_{2}}^{\mathrm{kSZ}}}{T}
      \contraction[2ex]{\langle T_{\bm{\ell}_{1}}^{\mathrm{kSZ}}}{T}{_{\bm{\ell}_{2}}^{\mathrm{kSZ}}T_{\bm{\ell}_{3}}^{\prime \mathrm{kSZ}}}{T}
      \langle T_{\bm{\ell}_{1}}^{\mathrm{kSZ}}T_{\bm{\ell}_{2}}^{\mathrm{kSZ}}T_{\bm{\ell}_{3}}^{\prime \mathrm{kSZ}}T_{\bm{\ell}_{4}}^{\prime \mathrm{kSZ}} \rangle} \rightarrow \mathrm{(d2)}\\
    +&\ {\contraction{\langle}{T}{_{\bm{\ell}_{1}}^{\mathrm{kSZ}}T_{\bm{\ell}_{2}}^{\mathrm{kSZ}}T_{\bm{\ell}_{3}}^{\prime \mathrm{kSZ}}}{T}
      \contraction[2ex]{\langle T_{\bm{\ell}_{1}}^{\mathrm{kSZ}}}{T}{_{\bm{\ell}_{2}}^{\mathrm{kSZ}}}{T}
      \langle T_{\bm{\ell}_{1}}^{\mathrm{kSZ}}T_{\bm{\ell}_{2}}^{\mathrm{kSZ}}T_{\bm{\ell}_{3}}^{\prime \mathrm{kSZ}}T_{\bm{\ell}_{4}}^{\prime \mathrm{kSZ}} \rangle} \rightarrow \mathrm{(d3)}\\
    +&\ {\langle T_{\bm{\ell}_{1}}^{\mathrm{kSZ}}T_{\bm{\ell}_{2}}^{\mathrm{kSZ}}T_{\bm{\ell}_{3}}^{\prime \mathrm{kSZ}}T_{\bm{\ell}_{4}}^{\prime \mathrm{kSZ}} \rangle}_{\mathrm{c}} \rightarrow \mathrm{(d4)}
  \end{aligned}.
\end{equation}

\(\mathrm{(b1)}\), \(\mathrm{(c1)}\), \(\mathrm{(d1)}\), \(\mathrm{(d2)}\), \(\mathrm{(d3)}\) are disconnected terms of Eq.~\eqref{eq:trispectrum}, which should be accounted for in the reconstruction Gaussian bias in Eq.~\eqref{eq:noise substraction}. Since \(T^{\mathrm{kSZ,g}}\) produces the same terms, these terms cancel in Eq.~\eqref{eq:ksz bias}.

\(\mathrm{(a1)}\) is a connected term of Eq.~\eqref{eq:trispectrum}, which includes a $<T^{\mathrm{kSZ}}T^{\mathrm{kSZ}}\phi>$ bispectrum. \(\mathrm{(b2)}\), \(\mathrm{(b3)}\), \(\mathrm{(c2)}\), \(\mathrm{(c3)}\) are connected terms of Eq.~\eqref{eq:trispectrum}, which include kSZ-lensing two-point coupling introduced in \cite{cooray03_weak_lensin_cmb, Kesden:2002jw}.
All these connected terms arise from the kSZ-lensing correlation, and are only non-negligible for the case of late-time kSZ~\cite{ferraro18_bias_to_cmb_lensin_recon}.  They do not cancel in  Eq.~\eqref{eq:ksz bias} by \(T^{\mathrm{kSZ,g}}\), and contribute to the bias.

\(\mathrm{(b4)}\) and \(\mathrm{(c4)}\) are connected terms of Eq.~\eqref{eq:trispectrum}. They include the connected part of the 4-point function involving two kSZ and two lensing fields (denoted by the subscript c), which do not cancel in Eq.~\eqref{eq:ksz bias}. These terms may arise from both the intrinsic non-Gaussianity of \(\phi\) and \(T^{\mathrm{kSZ}}\).

\(\mathrm{(d4)}\) is a connected term of Eq.~\eqref{eq:trispectrum}, which includes the connected part of kSZ trispectrum due to its non-Gaussianity. It does not cancel in
Eq.~\eqref{eq:ksz bias}, and exists for both late-time kSZ and reionization kSZ. To estimate the contribution of this term, we can run CMB lensing reconstruction algorithm on $T^{\mathrm{kSZ}}$ as Eq.~\eqref{eq:ksz trispectrum bias} shows.

\section{The Kinematic SZ Effect}
\label{sec:appendix B}
The temperature fluctuations induced by the kSZ effect in a direction \(\hat{\mathbf{n}}\) are given by (in units with \(c=1\))
\begin{equation}
  \label{eq:ksz1}
    \frac{\Delta T^{\mathrm{kSZ}}(\hat{\mathbf{n}})}{T_{\mathrm{CMB}}} = -\int d \eta g(\eta) \mathbf{p}_{e} \cdot \hat{\mathbf{n}}\ ,
\end{equation}
where \(\eta(z)\) is the comoving distance to redshift \(z\), \(g(\eta)\) is the visibility function, and \(\mathbf{p}_{e}\) is the peculiar electron momentum.

The visibility function \(g(\eta)\) represents the Poissonian probability that a photon is
last scattered at a time \(\eta\) which can be given by \(g(\eta)=(d\tau/d\eta)e^{-\tau}=\sigma_{T}n_{e,0}ae^{-\tau}\).  \(n_{e,0}\) is mean physical number density of free electrons, \(a\) is the scale factor and \(\tau\) is the optical depth. We  define  \(\mathbf{p}_{e}=(1+\delta_{e})\mathbf{v}_{e}\), where \(\delta_{e}\) is the free electron density contrast, and $\mathbf{v}_{e}$ is the electron velocity. The free electron density contrast can be expressed as ${\delta}_{e}= (1+\delta)(1+\delta_{x})$, where $\delta$ is the gas density contrast and \(\delta_{x}\) is the ionization contrast. So we have
\begin{equation}
  \label{eq:ksz3}
  \frac{\Delta T^{\mathrm{kSZ}}(\hat{\mathbf{n}})}{T_{\mathrm{CMB}}} = -\sigma_{T} \int \frac{d \eta}{1+z} e^{-\tau}(1+\delta+\delta_{x}+\delta\delta_{x}) \mathbf{v}_{e} \cdot \hat{\mathbf{n}},
\end{equation}
where the first contribution ($\propto\mathbf{v}_{e}$) is referred to as the ``Doppler'' term, the second contribution ($\propto\delta \mathbf{v}_{e}$) is the ``Ostriker-Vishniac'' term, and the third contribution ($\propto\delta_{x} \mathbf{v}_{e}$) is the ``patchy'' term.

 The ionization is inhomogeneous during the epoch of reionization, i.e., \(\delta_{x} \neq 0\). The modeling of reionization kSZ used in this work is described in \cite{McQuinn_2005, Alvarez:2015xzu}.

\bibliography{msm.bib}
\end{document}